# SCIENTIFIC REPORTS

natureresearch

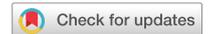

**OPEN** 

# A synergic approach to enhance long-term culture and manipulation of MiaPaCa-2 pancreatic cancer spheroids

Marta Cavo[1], Donatella Delle Cave[2], Eliana D'Amone[1], Giuseppe Gigli[1,3], Enza Lonardo[2] & Loretta L. del Mercato[1] ✉

Tumour spheroids have the potential to be used as preclinical chemo-sensitivity assays. However, the production of three-dimensional (3D) tumour spheroids remains challenging as not all tumour cell lines form spheroids with regular morphologies and spheroid transfer often induces disaggregation. In the field of pancreatic cancer, the MiaPaCa-2 cell line is an interesting model for research but it is known for its difficulty to form stable spheroids; also, when formed, spheroids from this cell line are weak and arduous to manage and to harvest for further analyses such as multiple staining and imaging. In this work, we compared different methods (i.e. hanging drop, round-bottom wells and Matrigel embedding, each of them with or without methylcellulose in the media) to evaluate which one allowed to better overpass these limitations. Morphometric analysis indicated that hanging drop in presence of methylcellulose leaded to well-organized spheroids; interestingly, quantitative PCR (qPCR) analysis reflected the morphometric characterization, indicating that same spheroids expressed the highest values of CD44, VIMENTIN, TGF-β1 and Ki-67. In addition, we investigated the generation of MiaPaCa-2 spheroids when cultured on substrates of different hydrophobicity, in order to minimize the area in contact with the culture media and to further improve spheroid formation.

Solid tumours grow in a three-dimensional (3D) conformation that expose cells to very specific conditions, such as hypoxia and a heterogeneous distribution of nutrient levels, all characteristics that affect cell fate[1]. In conventional monolayer cell cultures most of these environmental cues are missing and, consequently, bi-dimensional (2D) systems are often related to unsuccessful and contradictory results[2]. Contrariwise, 3D cultures more accurately recapitulate *in vivo* cell-cell interactions, consequently the importance of the third dimension in the study of cancer pathogenesis and evolution, as well as in the evaluation of drug efficacy, has been widely recognized by the scientific community[3–8]. With this in mind, many types of *in vitro* 3D cultures have been developed to recapitulate tumour growth conditions; one of the most common way to culture tumour cells in 3D is the spheroid model, where cells expand as spheres that reproduce cell-cell and cell-matrix interactions found in solid tumours[9]: in particular, the concentric arrangement and growth pattern of cells within spheroids mimic initial and avascular stages of solid tumours *in vivo*, not-yet vascularized micro-metastatic foci and hypoxic and necrotic regions far from the capillaries[10–12]. The first spheroid system was adapted to cancer research almost 40 years ago and significantly changed the landscape of preclinical studies, advancing our knowledge on cellular response to diverse therapeutic strategies, such as radiotherapy, hyperthermia, chemotherapy-based target-specific approaches and immunotherapy[13–15].

Among solid cancers, pancreatic ductal adenocarcinoma (PDAC) stands as the fourth leading cause of cancer-related death in the United States, with a 9% 5-year survival rate[16]; consequently, it is particularly urgent to overcome the current limitations providing models that replicate more precisely the biophysics of this tumour. The current goal of scientific community is going towards precision medicine approaches and consequently working with patients' derived cells, nevertheless cell lines are universally recognized as important models because they share characteristics with primary cells. In this regard, the MiaPaCa-2 pancreatic cancer cell line is widely used to study pancreatic ductal adenocarcinoma progression[17] because of several key factors: first of all it is KRAS mutant[18], making it an ideal target for anticancer drugs to develop proper therapeutic strategies;[19,20] secondly, even if they are epithelial cells they also show mesenchymal traits such as elevated expression of vimentin, a biomarker involved in the epithelial-to-mesenchymal transition (EMT)[21,22]. Through EMT cells become more motile and invasive and this process is involved in the metastasis initiation[23,24], thus targeting EMT has been one of the major challenges in cancer pharmacology[25]; lastly, MiaPaCa-2 cells are resistant at high doses of gemcitabine, the gold-standard drug for pancreatic cancer treatment[26]. Besides these characteristics, MiaPaCa-2 cell line form unstable and inhomogeneous sized spheroids that are often lost during harvesting and cell medium exchange, making hard to perform standard biological assays (such as quantitative PCR, western blots and flow cytometry) over time[27]. To be considered successful, spheroids should respect some specific criteria: diameter should reach at least 500 μm, in order to provide an external proliferating zone, an internal quiescent zone caused by lack of oxygen and nutrients and a necrotic core resembling the cellular heterogeneity of *in vivo* tumours[28]; secondly, spheroid culture should not only respect some morphogenic features but also maintain a functional activity and gene expression patterns[10].

[1]Institute of Nanotechnology, National Research Council (CNR-NANOTEC), c/o Campus Ecotekne, via Monteroni, 73100, Lecce, Italy. [2]Institute of Genetics and Biophysics "A. Buzzati-Traverso", National Research Council (CNR-IGB), Via Pietro Castellino 111, 80131, Naples, Italy. [3]Department of Mathematics and Physics "Ennio De Giorgi", University of Salento, via Arnesano, 73100, Lecce, Italy. ✉e-mail: loretta.delmercato@nanotec.cnr.it

Traditional methods for generating spheroids include hanging drop[29], spinner flasks[30], rotary cell culture systems[31], low binding and round bottom well plates[32] and hydrogel-based cultures[33]. Some of these methods have been adopted to generate spheroids with MiaPaCa-2 cells, but results have never been successful. In 2002, Sipos et al. generated spheroids from several PDAC cell lines except for MiaPaCa-2, as spheroids disaggregated during harvesting; authors defined this cell line "completely failing in growing as spheroids"[34]. In 2013, Yeon et al. affirmed that MiaPaCa-2 cells did not form spheroids and loose aggregates formed were not stable enough for harvest and further experiments[35]. Also when formed, as reported by Wen et al.[36], MiaPaCa-2 spheroids were analysed without being moved from their original plate and they were harvested just before dissociation: authors did not show any data about collecting spheroids while maintaining their integrity and did not perform any staining to characterize them. In 2016, Ware et al. generated spheroids using for the first time a novel approach that combined two traditionally used methods: the hanging drop and the use of methylcellulose (MC) in the media; this approach allowed increasing spheroid density, limiting their disaggregation. However, also in this case MiaPaCa-2 cells generated less dense and less circular spheroids compared to other pancreatic cell lines like BxPC-3 and Capan-1[10]. Therefore, despite the evident scientific need of using MiaPaCa-2 cell line in studying pancreatic cancer disease, generation and manipulation of homogeneous and stable MiaPaCa-2 spheroids over time remains a main challenge today, generating a trend to use other cell lines.

In this work, we compared different methods to produce spheroids with MiaPaCa-2 cell line in order to achieve the best possible spheroid configuration. This was a two-steps process: first of all, we compared six different techniques (i.e. hanging drop, round-bottom wells and Matrigel embedding, each of them with or without methylcellulose in according to recent publications of other authors) to evaluate which one was ideal for generating stable spheroids. Assessed that hanging drop plus methylcellulose was the best method, leading to compact and stable spheroids, we investigated the possibility to further improve the formation of MiaPaCa-2 spheroids by changing the substrates used for generation. It is known that hydrophobic surfaces minimize the area in contact with the culture media and limit cell location in the z-axis, forcing them all into the same plane. For this reason, we tested the performances of spheroids obtained by hanging drop plus methylcellulose on three substrates of different hydrophobicity (i.e. glass, standard Petri lids and PDMS). Spheroids were firstly characterized from a morphometric point of view, including dimension and aspect ratio parameters; then, qPCR analyses of cancer stem cell-related genes (i.e. CD44), mesenchymal gene (i.e. VIMENTIN), cytokine gene (i.e. TGF-β1) and proliferation gene (i.e. Ki-67) were carried out on cells constituting spheroids in order to characterize them from a molecular point of view.

## Results

**Spheroid formation and culture.** In order to establish the best method to obtain MiaPaCa-2 spheroids with a clinically-relevant diameter (over 500 μm), we grew 3D colonies by different techniques. Depending on the used method we obtained tumour spheroids that differed for morphology and size. Among the tested methods, hanging drop-based spheroids (with and without MC) and round bottom well assay (only with MC-enriched media) promoted the spontaneous formation of cell aggregates (Fig. 1, panels A, B and D). During the first 10 days of culture a decrease in the size was observed for these categories: aggregates became more compact, dense and dark, forming solid spheroids; according to Zanoni *et al.* this time interval is called "spheroidization time" and it is characterized by strong cell-cell adhesions[37]. After this period, hanging drop-based spheroids increased in size because of cell proliferation, while in round bottom wells cell colonies tend to disaggregate. On the other side, Matrigel allowed the growth of dense and dark cellular colonies formed by the proliferation of single or small groups of cells (Fig. 1, panel E).

By comparing the different methods, we could define the growth of spheroids in Matrigel as "bottom up" (from single cells to the spheroid), while in hanging drops and round bottom wells spheroids comes by cellular compaction and therefore follow a "top down" approach. The only category that did not produce any spheroid or aggregate was the round bottom well plates without MC in the media (Fig. 1, panel C). During the first 3 days of culture, hanging drop spheroids were very weak and media changing caused their complete disaggregation (Fig. 2, panel A); trying to preserve spheroid integrity, we found as good compromise to replace just half of media (i.e. 10 μl) every 2 days (Fig. 2, panel B). As a result, spheroids obtained with hanging drop method underwent disintegration only at edge level (Fig. 2, panels C,D); this phenomenon was attenuated in the presence of MC, with more compact and solid spheroids also at edge levels (Fig. 2, panels E,F).

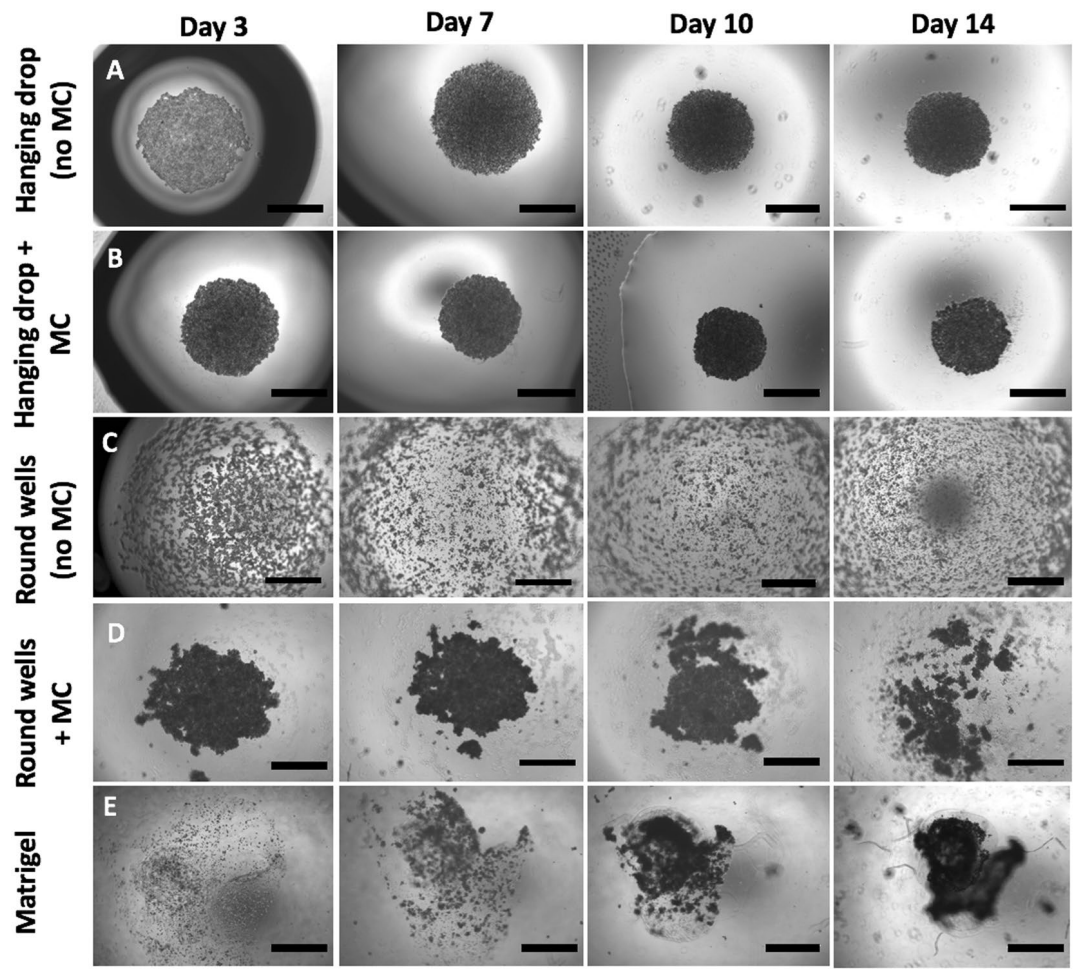

**Figure 1.** MiaPaCa-2 spheroids growth using different methods. The hanging drop and the round bottom well methods were tested both with simple media (**A,C**) and with MC-enriched media (**B,D**) to increase media viscosity and thus to improve cellular aggregation. Cell-embedded Matrigel method was carried out in standard 96 well-plates without the addition of MC (**E**). Spheroids were made of 30.000 cells per spheroid. Assembly time was 14 days for all methods. Bars = 1000 μm.

**Spheroid morphometric analysis.** To monitor the growth of spheroids obtained by using different approaches, different parameters, including the diameter and the aspect ratio, were investigated. The aspect ratio of a spheroid is defined as the ratio between the principal axis of revolution and the maximum diameter perpendicular to this axis[38]; values of aspect ratio increase with circularity: from 0 to the maximum value of 1.

Hanging drop-based spheroids followed a typical trend that foresees a first phase of reduction in diameter (until day 10), where cells are compacting, followed by a second phase of growth, where cells are proliferating (from day 10 to 14). As mentioned above, this is a typical phenomenon occurring in spheroids grown with hanging drop method, since first cells need to compact during the "spheroidization time" and then, only once strong cell-cell adhesions happened, they start to proliferate, increasing spheroid size. Contrariwise, aggregates produced in Matrigel continued to decrease in diameter. In round bottom plates with simple media, cells were not able to compact, thus values of diameter could not be calculated; in presence of MC, aggregates started to form but they completely lost their shape after 14 days. Notably, for each analysed category, we found spheroids with values of diameter much higher than what is usually reported in the literature: the ideal spheroid diameter is widely discussed, but no agreement has been reached yet. Small spheroids (<200 μm) are inappropriate for creating a physiological condition, since they do not develop chemical or proliferative gradients[39–42]. On the other side, large spheroids (>500 μm) will contain a hypoxic or necrotic core (desirable for drug testing) and cells will be in different proliferation stages, conferring heterogeneity to them[43]. In this work, all spheroids had a diameter bigger than 1 mm (Fig. 3, panel A); this could be caused by the used cell density (1.5 Millions/mL), suitable to obtain the most compact spheroids with MiaPaCa-2 cell line among the concentrations tested by us (data not shown); however, we do not consider this as a problem since spheroids resulted to be functionally active. The aspect ratio analysis showed that spheroids obtained with hanging drop methods, with or without MC, show the best circularity for all the time points, i.e. day 3, 5, 7, 10 and 14 (Fig. 3, panel B). Besides the morphometric analysis, we investigated the percentage of spheroids able to reach the end of the planned culture time (14 days) without disaggregating and meanwhile maintaining an aspect ratio equal or higher than 0.95. The hanging drop method resulted to be the most successful technique. Moreover, adding MC to media significantly improved the quality of spheroids, allowing to preserve 50% of them instead of 25% without MC (Fig. 3, panel C). Figure 3, panel D shows the reduction in diameter occurring in hanging drop-based spheroids in presence of MC (compared to the same method without MC).

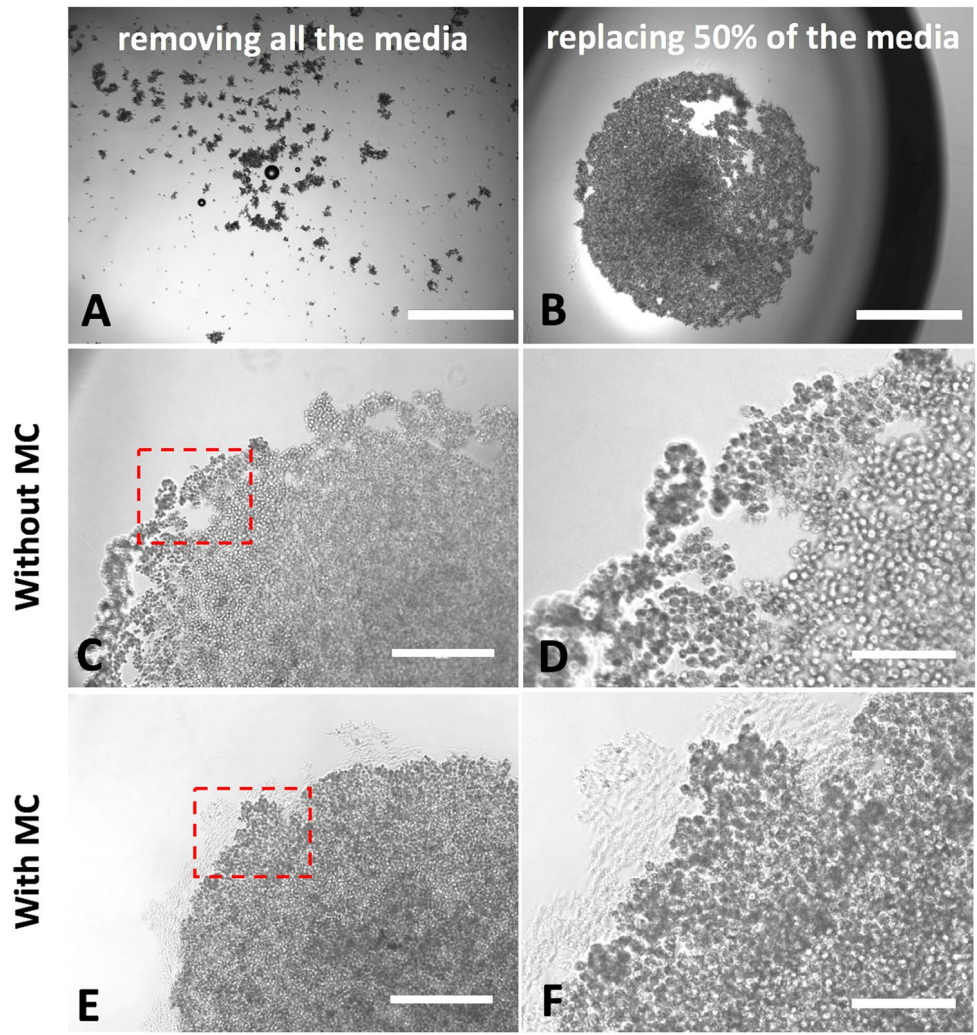

**Figure 2.** Effect of complete (**A**) and partial (**B**) media changing on spheroids. Without MC, the partial media changing allowed to preserve spheroids, but they disaggregated at edge level (panel C); this did not occur in presence of MC in the media (panel E). Panels D, F are zoomed areas of dashed boxes in panels C and E, respectively. Bars in panels A, B: 1000 µm; bars in panels C, E: 400 µm; bars in panels D, F: 200 µm.

**Spheroid handling and cellular viability.**  To further validate our hanging drop protocols, the live/dead cytotoxicity assay based on PI and Calcein AM staining was integrated with CLSM analyses. Here, the PI staining indicates compromised cell membranes with subsequent binding to intracellular nucleic acids (red fluorescence), whereas calcein AM fluorescence shows metabolically viable cells (green fluorescence). Representative CLSM images of spheroids grown with or without MC and stained for live/dead cytotoxicity assay are shown in Fig. 4. Analysis of CLSM images demonstrates that spheroids grown without MC disaggregated during multiple staining/washing steps (Fig. 4, Panel A), while spheroids grown with MC did not disassemble (Fig. 4, Panels C and E), highlighting that the proposed method with MC generates spheroids that can be efficiently handled during multi-step processing such as staining. Notably, by the end of 14 days, live/dead markers indicated that, unlike spheroids grown with MC were about 1 mm in diameter, the majority of cells remained viable both in external and internal regions (Fig. 4, panels D and F, respectively). From the histogram resulting from the quantification analysis, the percentage of dead cells on total amount of cells was calculated, finally providing a value ranging from ~5,5% to ~8% (n = 3), independently on the culture strategy adopted.

**Analysis of hanging drop spheroids grown on different substrates.**  The influence of surface hydrophobicity on the growth spheroids was analysed by culturing the hanging drop-based spheroids on three different substrates: glass coverslip, standard Petri lid and PDMS monolayer. In particular, due to its structure PDMS is known to be hydrophobic in nature[44].

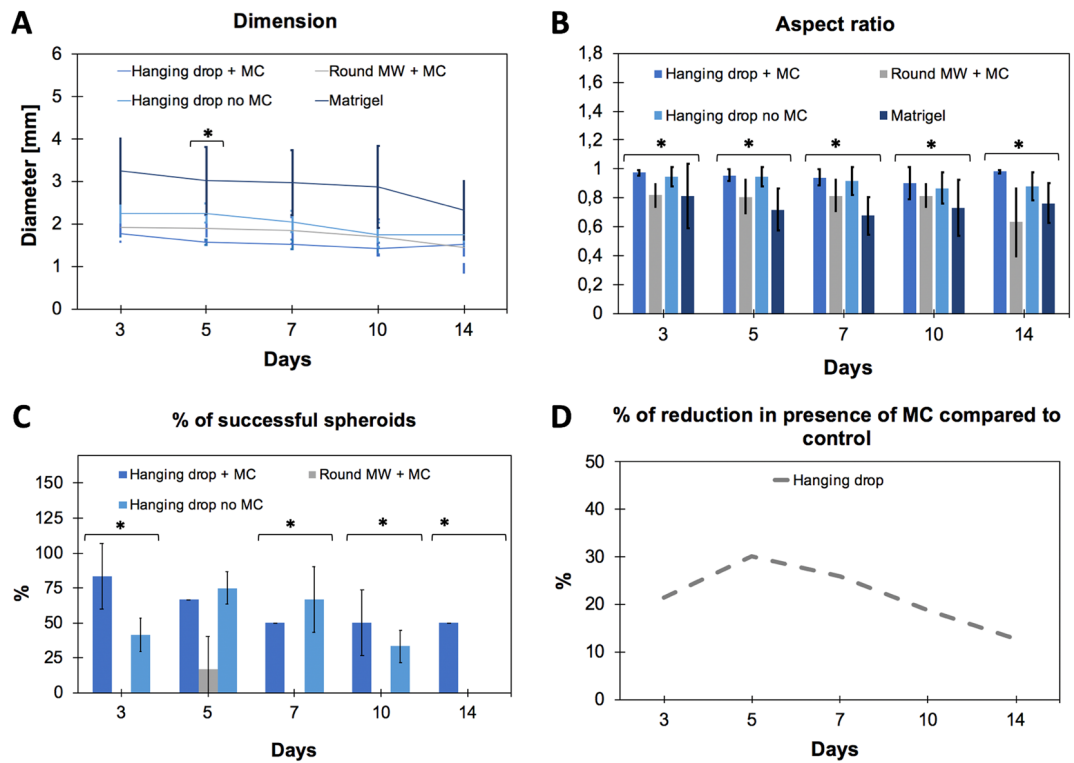

**Figure 3.** Spheroid growth characterization. Evaluation of diameter (**A**), aspect ratio (**B**) and percentage of success (percentage of spheroids able to reach the end of culture while maintaining an aspect ratio value $\geq$ 0.95) (**C**) of the different categories. Panel (D) shows the decrease in diameter of hanging drop spheroids when methylcellulose is used. In panels A, B and C data represent the mean ± standard error of the mean (SEM) (n = 20, for each category of spheroids). One-way between-group ANOVA test (p < 0.10) was used to determine the statistical differences between the different categories for each time point.

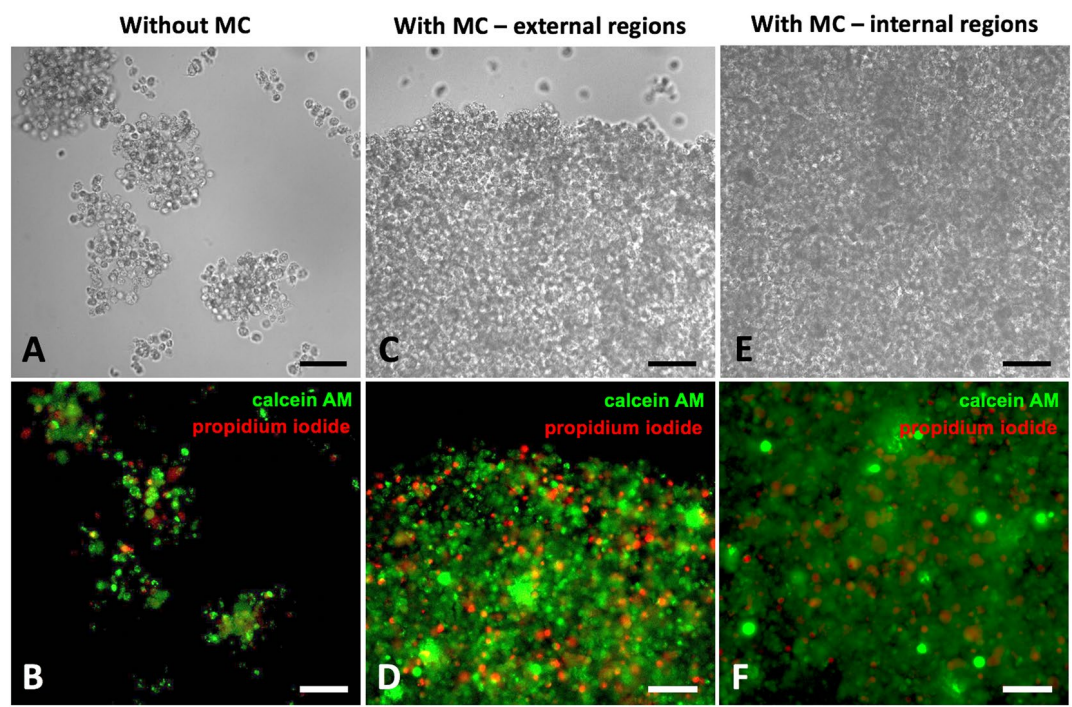

**Figure 4.** Live/dead assay on spheroids grown with hanging drop. Representative CLSM images of hanging drop-based spheroids without MC (**A,B**) and with MC (**C,D** external regions; **E,F** internal regions) stained for live (calcein AM; 3 µM solution; in green) and dead (propidium iodide; 10 µM solution; in red) cells after 14 days. Bars = 100 µm.

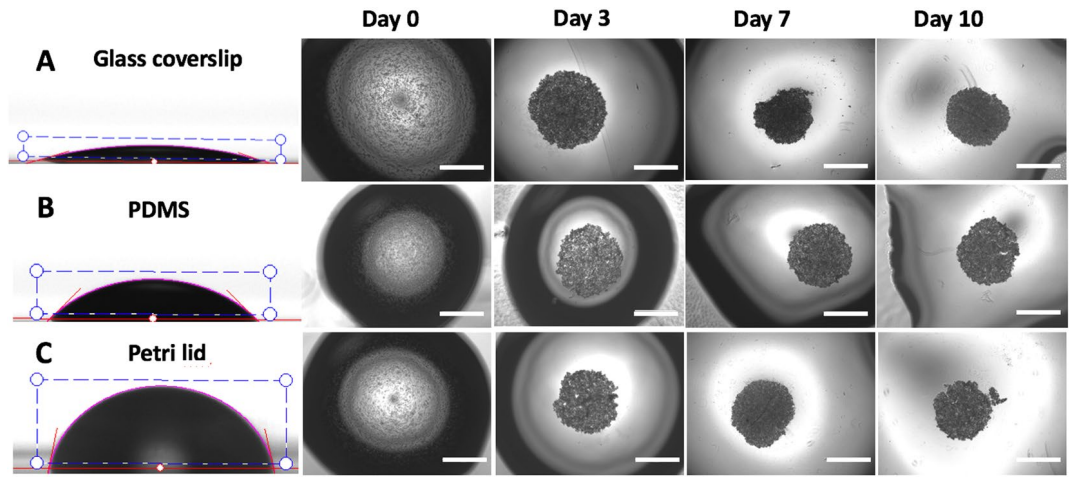

**Figure 5.** Development of hanging drop-based spheroids on substrate with different hydrophobicity. Drop of 20 μl of media containing 30.000 cells were placed on (**A**) glass coverslip, (**B**) PDMS monolayer and (**C**) standard Petri lid. Spheroids grown on these substrates were monitored up to 10 days. Bars = 1000 μm (4X objective lens).

The result of surface modification was evaluated by using contact angle measurements. PDMS is inherently hydrophobic[45], therefore PDMS coating significantly increased the surface hydrophobicity of glass slides. Glass slides had a contact angle of 15,77° ± 1,89°, PDMS slides of 42,93° ± 2,31° and Petri lids were the most hydrophobic with a contact angle of 78,51° ± 0,82° (Fig. 5, panels A, B, C).

This wide range of contact angles allowed exploring a hydrophobicity-related spheroid growth. Spheroids grown on these substrates were monitored up to 10 days in culture.

The diameter of each dispensed drop was strictly related to the substrate used (more hydrophobic the surface, smaller the diameter), thereby generating a more concentrated cell suspension when in contact with more hydrophobic surfaces (see cartoon in Fig. 6, panel A).

Morphometric analysis of spheroids included the evaluation of diameter and of aspect ratio (Fig. 6, panels A and B). According to what observed and reported, drops cultured on more hydrophobic substrates showed smaller diameter (related to higher compactness), while no significant differences were observed in aspect ratio values. The percentage of spheroids able to reach the 10th day of culture without disaggregating and meanwhile maintaining an aspect ratio equal or higher than 0.95 was also related to substrate hydrophobicity: indeed, spheroids cultured on glass coverslips were not able to be maintained in culture after 5 days (Fig. 6, panel D). More in detail, at days 7 and 10, among the spheroids grown on glass, none of those analysed had an aspect ratio at least equal to 0.95 (albeit very close to this value) and therefore the % of successful spheroids was evaluated as equal to zero. On the contrary, among the spheroids grown on PDMS and Petri lid some (or all) had an aspect ratio higher than or equal to 0.95, thus contributing to raising the% of successful spheroids.

**Spheroids gene expression analysis.** Cellular heterogeneity can be measured in several different ways, most commonly via genomic, epigenomic, transcriptomic, and proteomic studies. Here, we assessed the transcriptomic profile of our spheroids by analysing some key genes involved in the EMT/migration process (VIMENTIN), cancer stem cells phenotype (CD44), cytokine secretion (TGF-β1) and proliferation (Ki67), respectively. By qPCR analyses we characterized molecularly the spheroids cultured with hanging drop, hanging drop + MC and cell-embedded Matrigel protocols (Fig. 7). We observed an increase in CD44, VIMENTIN, TGF-β1 and Ki-67 in both hanging drop and hanging drop + MC methods compared to Matrigel. Interestingly, the greater differences in the gene expression for TGF-β1 and Ki-67 were reached after 14 days of culture, suggesting a time-dependent increasing in mesenchymal traits and proliferation, respectively. Of note, all the methods used to generate the 3D spheroids induced a marked increase in the expression levels of the analysed genes respect to the cells grown in 2D (ADH), thus, the behaviour of 3D-cultured cells is more reflective of *in vivo* cellular responses.

## Discussion

In recent years, tumour spheroids have gained great attention due to the limitations of monolayer cell cultures to precisely mimic *in vivo* structure and cellular interactions, increasing the use of these systems for basic research, drug screening and preclinical studies;[43,46,47] in parallel, many efforts have been made to develop new technologies to characterize the complex 3D organization of spheroids, especially because their difficulties in handling and carrying out analytical measurements[48]. Among solid tumours, pancreatic ductal adenocarcinoma (PDAC) remain a leading cause of death[16], also because the lack of proper preclinical models. In this work, we decided to compare different methods to generate tumour spheroids using MiaPaCa-2 cell line, an interesting model and an ideal target for anticancer drugs, but unable to form stable and robust spheroids to be harvested and used for complex analyses such as staining and live-cell imaging. To produce MiaPaCa-2-based spheroids, we played with three parameters, specifically: (i) different methods of preparation; (ii) different media compositions and (iii) substrates of different hydrophobicity.

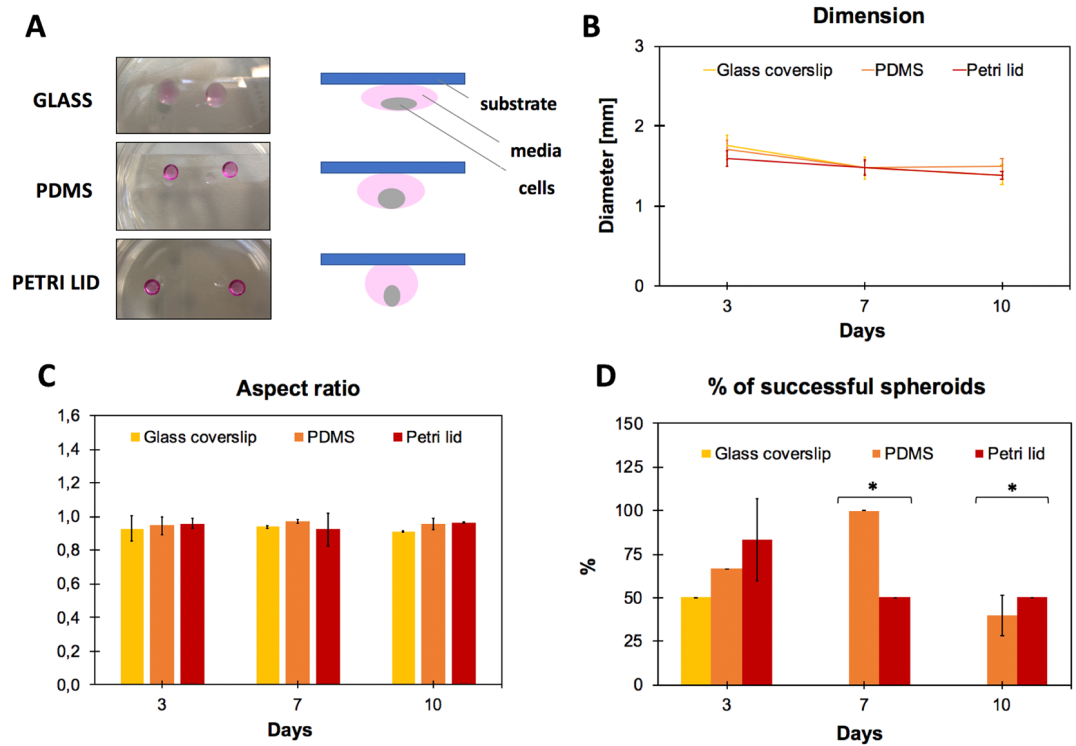

**Figure 6.** Spheroid growth characterization in relation to the substrate used. (**A**) Pictures and schematic representation of hanging drop cultures performed on glass, PDMS and Petri lid substrates. (**B,C,D**): analysis of diameter, aspect ratio and % of successful spheroids at day 10, respectively. One-way between-group ANOVA test ($p < 0.10$) was used to determine the statistical differences between the different categories for each time point.

First of all, we found that distinct differences in spheroid morphology were evident for each examined method: while the spheroids generated from hanging drop method with MC-enriched media could be individually transferred without major disruption of their architecture and be processed for further analyses (e.g. live/dead staining), inconsistencies in compaction were present for all the other typologies. Spheroid compaction was observed from the decrease in spheroid diameter accompanied by an increase in darkness and cell density; after this phase, spheroids started to proliferate increasing in diameter. Then, assessed that hanging drop plus methylcellulose leaded to compact and stable spheroids, we investigated the possibility to further improve the generation of MiaPaCa-2 spheroids by minimizing the area in contact with the culture media using hydrophobic surfaces. In this case, we found that the diameter of each dispensed drop was strictly related to the substrate used (more hydrophobic the surface, smaller the diameter), thereby generating a more concentrated cell suspension when in contact with more hydrophobic surfaces. Moreover, the percentage of spheroids able to reach the 10th day of culture without disaggregating and meanwhile maintaining an aspect ratio equal or higher than 0.95 was also related to substrate hydrophobicity: indeed, spheroids cultured on glass coverslips were not able to be maintained in culture after 5 days.

To provide a more complete picture of spheroid cellular situation, we performed a qPCR analysis to observe the presence of Cancer Stem Cells (CSCs) as tumour-initiating cells, which are key drivers in tumour progression, resistance and relapse. In particular, the CSCs are a subpopulation of cancer cells that share similar properties with normal stem cells, including the ability to self-renew and to differentiate in different cell types potentially able to reconstitute the tumour bulk. Albeit no specific markers have been identified to define the CSC sub-population, the expression of the CD44 gene is generally associated with CSC features[49]. qPCR analysis after 3, 7 and 14 days revealed considerably higher expressions of cancer stem cell-related gene CD44 in spheroids respect to adherent cells and this effect is enhanced in those generated in hanging drop with MC-enriched media. Moreover, a hallmark of cancers is their ability to invade the extracellular matrix and reach distant organs, thus leading to metastases. In this regard, the Epithelial-to-Mesenchymal Transition (EMT) is the driving force which allows cells to achieve a motile and invasive phenotype[50] and the main feature of EMT is the loss of epithelial characteristics and the acquisition of a mesenchymal phenotype such as an increase in VIMENTIN expression. Again, it has been widely established that the TGF-β1, a key member of the TGF-β superfamily, has a crucial role in EMT induction[51]. Notably, qPCR analysis after 3, 7 and 14 days revealed that spheroids dramatically upregulated the expression of the EMT markers (VIMENTIN and TGF-β1) respect to 2D cultures. Finally, the expression of the proliferative gene Ki-67 is also upregulated in 3D cultures compared to 2D; specifically, the expression of the Ki-67 is strictly associated with cell proliferation and in patients is often correlated with the clinical course of the disease. The high level of Ki-67 in these spheroids is thus a crucial parameter, meaning that – besides the huge dimension of spheroids – cells are maintaining their proliferation activity. Most of these genes showed a higher level of expression in hanging drop-based spheroids than in Matrigel-based ones. These data indicate that RNA analyses are positively correlated to the morphometric analysis results. Taken together, these results appear of great interest for at least two main reasons: firstly, they demonstrate that a combinational approach, given by the use of MC-enriched media and a properly hydrophobic substrate, may lead to the generation of robust MiaPaCa-2 spheroids with a critical volume that can be harvested and used for further analysis such as staining and confocal imaging; secondly, the coherence between morphometric and genetic analyses suggest that spheroid's shape and volume may reflect a different genetic situation of cells composing the spheroids. Overall, we think that the methods presented here will provide new tips for optimizing the long-term culture and manipulation of human pancreatic tumours cells characterized by low cohesiveness and low manageability. Future research in our laboratories will be focused on applying our findings using more appropriate cell line models[52], including primary cells from patients with a known mutational profile, to grow spheroids with improved homogeneity and stability and to evaluate their drug-response.

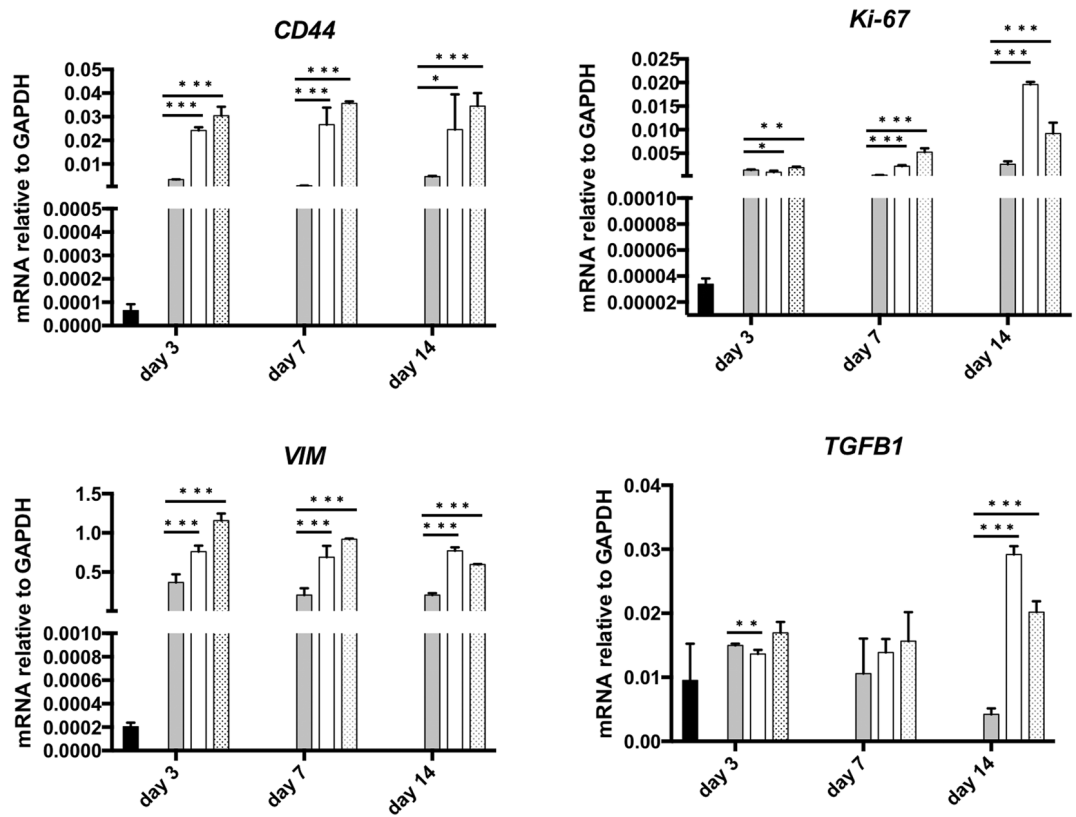

**Figure 7.** qPCR analysis. Analysis of CD44, Ki-67, Vimentin and TGF-β1 genes in spheroids cultured in adherent conditions (ADH), Matrigel, hanging drop and with MC-enriched media (HD + MC). Data are normalized to GAPDH expression and are represented by the mean ± SEM (n > 3). Multiple t-test was used to compare the result in pairs (***p < 0.0005, **p < 0.005, *p < 0.05).

## Methods

All reagents were purchased from Sigma Aldrich (St Louis, MO, USA), unless specified otherwise.

**Cell culture.** Human pancreatic cancer cell line MiaPaCa-2 was obtained from ATCC (American Type Culture Collection, Rockville, Md., USA) and cultured in ATCC Dulbecco's Modified Eagle Medium (DMEM) supplemented with 10% fetal bovine serum (FBS), 2.5% horse serum and 1% antibiotics (penicillin-streptomycin). Subculture was carried out in Corning T-75 flasks and media were changed three times per week. Cells were cultured till ~70% confluency, trypsinized per regular passage and counted on a hemocytometer.

**Spheroid development.** Spheroids were created using different approaches: the hanging drop method, the round bottom wells and the cell-embedded Matrigel. The hanging drop and the round bottom well methods were tested with and without the use of MC in the medium. To prepare MC-enriched medium, complete medium was supplemented with 20% MC stock solution. For preparation of MC stock solution we followed the indications provided by other authors[10]: in detail, 6 g of autoclaved MC powder (M0512; Sigma-Aldrich) was dissolved in preheated 250 mL basal medium (60 °C) for 20 minutes. Then, 250 mL of medium (room temperature) containing double the amount of FBS for the particular cell line was added to a final volume of 500 mL and the whole solution was mixed overnight at 4 °C, as suggested by other authors. The final stock solution was aliquoted and cleared by centrifugation (5000 rpm for 2 h at room temperature). Only the clear, highly viscous supernatant was used for the spheroid formation, which was approximately 90–95% of the stock solution. After these passages, final MC concentration is about 0.24%.

| Gene symbol | Forward primer (5′->3′) | Reverse primer (5′->3′) |
|---|---|---|
| **CD44** | CACGTGGAATACACCTGCAA | GACAAGTTTTGGTGGCACG |
| **GAPDH** | CAGGAGCGAGATCCCT | GGTGCTAAGCAGTTGGT |
| **Ki-67** | GTGCAGAGAGTAACGCGG | ACACACATTGTCCTCAGCCTTC |
| **VIMENTIN** | TGCCCTTAAAGGAACCAATG | CTCAATGTCAAGGGCCATCT |
| **TGF-β1** | AAGTGGACATCAACGGGTTC | TGCGGAAGTCAATGTACAGC |

**Table 1.** Genes analysed and forward/reverse primer sequences used for qPCR analyses.

For all the experiments cells were suspended in media at a concentration of 1.5 millions/ml. For hanging drop, twenty microliter drops containing 30.000 cells were pipetted onto the lid of 100 mm dishes and were inverted over dishes containing 10 mL phosphate buffer solution to avoid drying. For round bottom wells, the same number of cells was seeded into wells of 96-well round-bottom plates and allowed to compact for few hours before adding supplementary media. For cell-embedded Matrigel assay, cells were suspended at a concentration of 1.5 millions/mL in cold Matrigel and pipetted into standard 96-well plates, allowed to gel in incubator for 45 minutes and then complete media was added.

All the cultures were incubated under standard conditions (5% $CO_2$, 37 °C) for 14 days. Three days following initial plating, media was changed for the first time and this procedure was carried out every two days; for hanging drop method, the partial (50%) replace of the media was preferred to the complete change to avoid spheroid disruption (see Fig. 2, panels A and B).

**Observation of spheroid formation and morphometry.** Plates were removed every day from incubator for imaging to monitor the formation of spheroids within each hanging drop or well. For each cell culture, 3–5 representative images were obtained by using a phase contrast microscope (EVOS XL Core Imaging System). The NIH ImageJ software (https://imagej.nih.gov/ij/) was used to measure the diameter and the aspect ratio of the spheroids. The percentage of successful spheroids was calculated as the quantity of spheroids able to reach the end of culture (14 days) without disaggregating and maintaining an aspect ratio value equal or higher to 0.95.

**Cellular viability in hanging drop spheroids.** Cellular viability in hanging drop spheroids (with and without MC) was evaluated at day 14. Calcein AM (live dye) was added to final concentration of 3 μM, while propidium iodide (PI, dead dye) was added to a final concentration of 10 μM to each hanging drop. Following a 45-minute incubation at 37 °C, hanging drops were harvested onto a pre-cleaned glass microscope slide and imaged using an upright confocal laser-scanning microscope (CLSM; SP8 Leica). Fluorescence images were obtained at every z-axis encompassing the spheroids, at 488 nm for Calcein-AM (live cells: green) and 561 nm for PI (dead cells: red). This staining/washing sequence was also used to evaluate the handling of spheroids at the end of the culture period. Projections of the z-stack images were obtained using ImageJ. In order to have a quantitative estimation of the percentage of dead cells, we performed a semi-automatic particle analysis. In detail, the images were first converted to "binary", black and white, images. A threshold range was set to tell the objects of interest (dead cells, stained by Propidium Iodide) apart from the background. All pixels in the image whose values lie under the threshold were automatically converted to black and all pixels with values above the threshold were converted to white. The resulting histogram was plotted.

**Preparation of substrates with different hydrophobicity for hanging drop.** Three different surfaces, namely glass, polydimethylsiloxane (PDMS) and standard Petri lid, were prepared to evaluate the impact of surface hydrophobicity on spheroid formation. The PDMS solution (SYLGARD 184, Dow Corning, USA) was prepared by mixing the elastomer and the curing agent with ratio of 10:1 (w/w), cast on glass slides as a thin uniform coating and baked at 140 °C for 15 minutes. The cross-linked PDMS-coated glasses (hereinafter referred to as PDMS slides) and standard glass microscope slides were sterilized under UV for 1 hour before use.

**Contact angle measurements.** Static water contact angles were measured by using the sessile drop method and a CAM 200 (KSV Instruments Ltd., Finland) instrument. The presented results correspond to an average of at least three measurements performed onto many areas of the samples.

**Growth of hanging drop spheroids on different substrates.** The PDMS and the glass slides were glued onto the bottom of Petri lids; for all the categories, hanging drop spheroids were created and grown as reported in section "Spheroid development". The characterization of spheroid diameter, aspect ratio and % of successful spheroids was evaluated up to day 10.

**Quantitative polymerase chain reaction (qPCR).** Total RNA from spheroids and adherent cells was extracted with TRIFAST (Euroclone) according to the manufacturer's instructions. One microgram of total RNA was used for cDNA synthesis with High-Capacity reverse transcriptase (Thermofisher). Then, qPCR was performed using SYBR Green master mix (Thermofisher), according to the manufacturer's instructions. The list of utilized primers is depicted in Table 1.

**Statistical analysis.** All experiments were repeated with at least three biological replicates with n ≥ 10, in order to carry out statistics. All data are represented as mean ± standard error of the mean (SEM). For morphometric characterization (Figs. 3 and 6), one-way between-group ANOVA test ($p < 0.10$) was used to determine the statistical differences between the different methods for each time point. For qPCR analysis (Fig. 7) multiple t-test was used to compare the result in pairs.

### Acknowledgements

Authors gratefully acknowledge the ERC project INTERCELLMED (grant number 759959) and the project "TECNOMED—Tecnopolo di Nanotecnologia e Fotonica per la Medicina di Precisione" [Ministry of University and Scientific Research (MIUR) Decreto Direttoriale n. 3449 del 4/12/2017, CUP B83B17000010001]. E.L. also acknowledges the financial support from the Marie Curie IF (H2020-MSCA-IF-2015, #703753), My First AIRC Grant (MFAG-2017, #20206) and POR Campania FESR 2014/2020 (Project SATIN).

### Author contributions
M.C. and L.L.D.M. conceived the idea. M.C. planned and carried out the experiments with the help from E.D.; D.D.C. and E.L. performed the gene expression analyses and analysed the results. M.C. wrote the paper with inputs from D.D.C., E.L., G.G. and L.L.D.M. All authors discussed the results and contributed to the manuscript. The entire research work was supervised by L.L.D.M.

### Competing interests
The authors declare no competing interests.

### Additional information
**Correspondence** and requests for materials should be addressed to L.L.d.M.

**Reprints and permissions information** is available at www.nature.com/reprints.

**Publisher's note** Springer Nature remains neutral with regard to jurisdictional claims in published maps and institutional affiliations.